\documentclass[twocolumn,prb,twocolumn,superscriptaddress]{revtex4-1}

\usepackage{amsmath,amssymb,mathrsfs}
\usepackage{natbib}
\usepackage{subfigure}
\usepackage{tabularx}
\usepackage{epsfig}
\usepackage{longtable}
\usepackage{amsfonts}
\usepackage{rotating}
\usepackage{subfigure}
\usepackage{amsmath}
\usepackage{comment}

\usepackage[unicode=true,bookmarks=true,bookmarksnumbered=false,bookmarksopen=false,breaklinks=false,pdfborder={0 0 1},backref=false,colorlinks=true]{hyperref}

\hypersetup{linkcolor=magenta,urlcolor=blue,citecolor=blue,pdfstartview={FitH},hyperfootnotes=false,unicode=true}

\def\be{\begin{equation}}
\def\ee{\end{equation}}
\def\bea{\begin{eqnarray}}
\def\eea{\end{eqnarray}}

\begin{document}

\title{Chiral Spin Condensation in a One-Dimensional Optical Lattice}

\author{Ying-Hai Wu}
\affiliation{Max-Planck-Institut f{\"u}r Quantenoptik, Hans-Kopfermann-Stra{\ss}e 1, 85748 Garching, Germany}

\author{Xiaopeng Li}
\email{xiaopeng\_li@fudan.edu.cn}
\affiliation{State Key Laboratory of Surface Physics, Institute of Nanoelectronics and Quantum Computing, and Department of Physics, Fudan University, Shanghai 200433, China}
\affiliation{Collaborative Innovation Center of Advanced Microstructures, Nanjing 210093, China} 
\affiliation{Condensed Matter Theory Center and Joint Quantum Institute, Department of Physics, University of Maryland, College Park, MD 20742-4111, USA} 

\author{S. Das Sarma}
\affiliation{Condensed Matter Theory Center and Joint Quantum Institute, Department of Physics, University of Maryland, College Park, MD 20742-4111, USA}

\begin{abstract}
We study a spinor (two-component) Bose gas confined in a one-dimensional double-valley optical lattice which has a double-well structure in momentum space. Based on field theory analysis, it is found that spinor bosons in the double-valley band may form a spin-charge mixed chiral spin quasicondensate under certain conditions. Our numerical calculations in a concrete $\pi$-flux triangular ladder system confirm the robustness of the chiral spin order against interactions and quantum fluctuations. This exotic atomic Bose-Einstein condensate exhibits spatially staggered spin loop currents without any charge dynamics despite the complete absence of spin-orbit coupling in the system, creating an interesting approach to atom spintronics. The entanglement entropy scaling allows us to extract conformal-field-theory central charge and establish the low-energy effective field theory for the chiral spin condensate as a two-component Luttinger liquid. Our predictions should be detectable in atomic experiments through spin-resolved time-of-flight techniques.
\end{abstract}

\maketitle

\section{Introduction} 

Ultracold atoms confined in optical lattices provide a fascinating synthetic platform for quantum simulations of various lattice Hamiltonians in a controllable fashion beyond what is possible in natural crystals~\cite{1998_Zoller_Jaksch_PRL,2002_Hofstetter_Cirac_PRL,2007_Lewenstein_AP,2008_Bloch_Dalibard_RMP,2010_Esslinger_CMP,2015_Lewenstein_RPP,2016_Li_Liu_RPP}. For example, synthesizing optical lattices with artificial gauge fields has attracted considerable recent research efforts, and has now become one of the most important developments in ultracold atomic physics~\cite{2011_Dalibard_Gerbier_RMP}. In particular, $\pi$ flux models, which are in general difficult to find in solid state materials, have been realized with atoms in shaken optical lattices~\cite{2013_Chin_NatPhys,2013_Sengstock_NatPhys}. The artificial-gauge-field quantum simulator has become a versatile ground for investigating exotic many-body physics such as bosonic chiral condensates~\cite{2013_Paramekanti_PRB,2015_Edmonds_EPL,2014_Zaletel_PRB,2016_Li_Liu_RPP} and fermonic quantum Hall  states~\cite{1980_Klitzing_PRL}. 

Another optical lattice based quantum simulator recently attracting interest is laser-assisted spin-orbit coupling (SOC) ~\cite{2005_Ruseckas_PRL,2005_Osterloh_Zoller_PRL,2009_Liu_PRL,2011_Lin_Spielman_Nature,2012_Pan_SOC,2012_Zhang_Zhai_PRL,2012_Cheuk_Zwierlein_PRL,2013_Galitski_NatReview,2013_Xu_You_PRA,2013_Anderson_Spielman_PRL,2015_Zhai_RPP,2016_Huang_Zhang_NatPhys,2016_Wu_Pan_Science,2017_Bloch_SOC}, aiming for exotic Rashba ring condensate of bosons~\cite{2015_Zhai_RPP} and symmetry-protected topological states of fermions~\cite{2010_Hasan_RMP,2016_Chiu_RMP}. For bosons, it has been demonstrated that one-dimensional SOC leads to crystalline condensates~\cite{2013_Galitski_NatReview,2014_Ji_Pan_NatPhys}, and a more recent experiment realizing  two-dimensional SOC has further advanced this subject~\cite{2016_Wu_Pan_Science}. For fermions, the SOC effects have been observed in the single-particle energy dispersion~\cite{2012_Zhang_Zhai_PRL,2012_Cheuk_Zwierlein_PRL,2016_Huang_Zhang_NatPhys}. 
The developments in SOC quantum simulators together with the recently demonstrated capability of measuring atomic spin currents~\cite{2016_Bloch_Schweizer_arXiv} are bridging the fields of spintronics and ultracold atomic physics. 
However, the experimentally realized SOCs are all single-particle effects extrinsically induced by laser, causing experimental challenges (e.g., heating problems), in studying many-body quantum effects with more sophisticated SOCs. It is thus worthwhile to find alternative ways to generate SOC-like effects, e.g. with interactions, so that nontrivial strong correlations and many-body effects can be studied. Previous mean field analysis shows that SOC effects can spontaneously emerge due to two-body interactions~\cite{2014_Li_NatComm}, but whether this exotic phenomenon could survive (beyond mean field theory) against strong fluctuations is unknown, raising possible experimental difficulties in realizing this unconventional paradigm for SOC engineering. 

\begin{figure}
\includegraphics[width=0.5\textwidth]{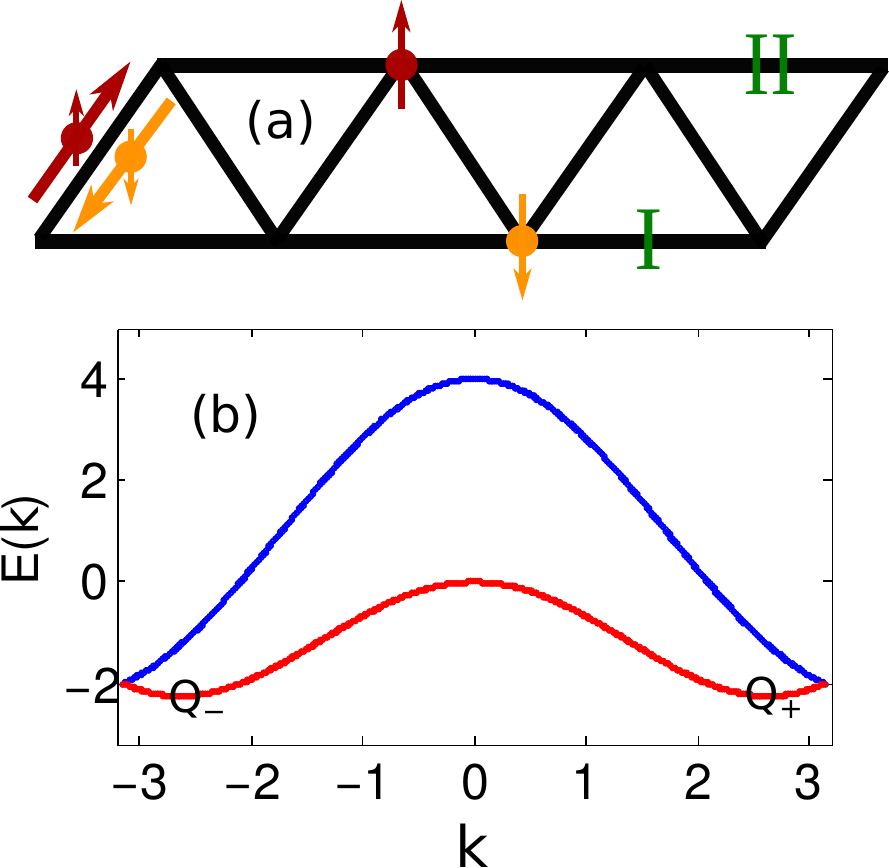}
\caption{{(a) Schematics of the $\pi$ flux triangular ladder model. The two legs are labeled as I and II. The two components are represented using up and down arrows. In the thermodynamic limit, the two components develop currents along opposite directions as indicated by the arrows along the bond. (b) The single-particle band structure with two valleys at momenta $Q_{\pm}$.}}
\label{Figure1}
\end{figure}

In this paper, we study a spinor (two-component) Bose gas in a one-dimensional double-valley lattice in the absence of any bare SOC. The double-valley lattice model describes the $sp$-orbital coupled optical lattice system~\cite{li2013topological} as realized in Ref.~\onlinecite{2013_Chin_NatPhys} or the shaking-induced two-dimensional $\pi$-flux triangular lattice~\cite{2013_Sengstock_NatPhys} reduced to one dimension. For certain repulsive spin-miscible interactions~\cite{pethick2002bose}, it is found through an effective field theory analysis that the ground state could be a chiral spin condensate in the presence of strong fluctuations in one dimension. This chiral spin condensate exhibits spin-charge mixing, which turns out to be the major ``quantum-fluctuation-source" in selecting the nontrivial chiral spin order in a classically degenerate manifold. For a specific $\pi$-flux triangular ladder model, we confirm the existence of chiral spin condensate using the density matrix renormalization group (DMRG) method~\cite{White1992,Schollwock2011}, which is a variational algorithm within the class of matrix product states and numerically ``exact" for one-dimensional systems due to their entanglement properties. This strongly correlated state features spontaneous staggered chiral spin loop currents where the two spin components counterflow, i.e. they move along opposite directions (Fig.~\ref{Figure1}). We emphasize that such spin-orbit effects arise purely from interactions in the theory, distinctive from previously explored single-particle SOC effects. By computing various correlation functions, conformal-field-theory central charge, and entanglement scaling, we conclude that the relevant low-energy effective field theory here is a two-component Luttinger liquid.

\section{Field Theory Analysis}

Consider two-component (pseudospin $\sigma=\uparrow,\downarrow$) bosons in a double-valley band, e.g. the experimentally realized one dimensional $sp$-orbital coupled optical lattice~\cite{2013_Chin_NatPhys} or the $\pi$-flux triangular lattice reduced to one dimension~\cite{2013_Sengstock_NatPhys}. Assuming the band minima are at ${\pm Q}$ (see Fig.~\ref{Figure1}), the low energy degrees of freedom $\phi_{\sigma,\pm}$, describing fluctuations near the band minima, are introduced as $\phi_{p = \pm, \sigma} (x) = \int \frac{dk}{2\pi}  b_{\sigma} (\pm Q + k) e^{ik x}$, with $b_{\sigma}(q)$ being the annihilation operator for the eigenmode of the lower band (Fig.~\ref{Figure1}). The effective field theory has the action $S = \int dx dt {\cal L}$ with the Lagrangian 
\bea 
{\cal L} &=& \phi_{p \sigma } ^* ( x,t) \left [ i\partial _t + \frac{1}{2M} \frac{\partial^2} {\partial x^2} - 
			i p\lambda \frac{\partial^3}{\partial x^3} + \mu   \right] \phi_{p \sigma }  (x,t )   \\ 
	&+& \sum_{\sigma \sigma' } g_{\sigma \sigma'}  
		\left [\sum_{p p' } |\phi_{p \sigma } |^2 | \phi_{p' \sigma } | ^2 
			+ \sum_p \phi_{p \sigma } ^* \phi_{-p \sigma} \phi_{-p \sigma'} ^* \phi_{p\sigma' }  \right ],  \nonumber 
\eea 
where the effective mass $M$ and the third order derivative term $\lambda$ can be extracted from the energy dispersion near the band minima, and $g_{\sigma\sigma'}$ characterizes the interaction strengths in the spinor Bose gas. We introduce 
$g_{\uparrow \uparrow} = g_{\downarrow \downarrow} = g_d$ 
and
$g_{\uparrow \downarrow} = g_{\downarrow \uparrow} = g_{od} $. 
For spin-miscible repulsive interactions, we have $g_d >0$, $g_{od} >0$, and $g_d ^2 > g_{od}^2$. Because of the exchange term $\sum_p \phi_{p \sigma } ^* \phi_{-p \sigma} \phi_{-p \sigma'} ^* \phi_{p\sigma' } $, we cannot have coexistence of $\phi_{+ \sigma}$ and $\phi_{-\sigma}$ at low energy. This leads to a spontaneous Ising symmetry breaking at the classical level which remains robust even in the presence of quantum fluctuations at zero temperature. We thus have either chiral charge or chiral spin superfluid  as characterized by 
\be
\left[ 
	\begin{array}{c} 
	\phi_{\uparrow +} \\ 
	\phi_{\downarrow+} 
	\end{array} 
\right] 
\text{or} 
\left[ 
	\begin{array}{c} 
	\phi_{\uparrow +} \\ 
	\phi_{\downarrow -} 
	\end{array} 
\right] 
=  
\left[ 
\begin{array}{c} 
 \sqrt{\rho_0 + \delta \rho_\uparrow } e^{i\theta_\uparrow} \\ 
 \sqrt{\rho_0 + \delta \rho_\downarrow}  e^{i\theta_\downarrow} 
\end{array} 
\right],   
\ee 
respectively, where the fields $\delta \rho_\sigma$ and $\theta_\sigma$ represent density and phase fluctuations at  low energy. 

We define charge and spin degrees of freedom as $ \theta_{c,s} = \frac{1}{\sqrt{2}} \left( \theta_\uparrow \pm \theta_\downarrow \right) $, whose conjugate momenta $\Pi_{c,s}$ are introduced correspondingly. For the chiral charge superfluid, we find a spin-charge separated Hamiltonian 
\bea 
{\cal H}_{\chi_c}  &=& \sum_{\nu = c,s}  \frac{1}{2} v_\nu \left[ K_\nu \Pi_\nu ^2 + \frac{1}{K_\nu} (\partial_x \theta_\nu)^2 \right] \nonumber \\ 
&& + \lambda \left[ \Pi_c \frac{\partial^3}{\partial x^3 } \theta_c + \Pi_s \frac{\partial^3}{\partial x^3 } \theta_s \right], 
\label{eq:chic} 
\eea 
with $K_{c/s} = \sqrt{2(g_d \pm g_{od} ) /\rho_0} $ 
and $v_{c/s}  = \sqrt{2 \rho_0 (g_d \pm g_{od}) } $. 
Here we have neglected higher order nonlinear terms as we shall show later that the classical degeneracy between chiral spin and charge superfluids is already broken at the harmonic order. Performing the expansion in terms of the eigenmodes for the quantum fields, 
\bea
\theta_{\nu} &=& \sqrt{K_\nu} \int_{-\Lambda} ^\Lambda \frac{dk}{2\pi} \frac{1}{\sqrt{2 |k|}} 
\left[		a_\nu (k) e^{ikx} + a_\nu ^\dag (k) e^{-ikx}  \right]  \\ 
\Pi_{\nu} &=& \frac{-i}{\sqrt{ K_\nu } } \int_{-\Lambda} ^\Lambda \frac{dk}{2\pi}  \sqrt{ \frac{|k|}{2} }  
	\left[  a_\nu (k) e^{ikx} - a_\nu ^\dag (k) e^{-ikx}   \right] 
\eea 
we find that the energy density of the ground state is $\frac{1}{4\pi} [v_c + v_s] \Lambda^2$, with $\Lambda$ the high-momentum cutoff in the field theory. Note that the third order derivative term $\lambda$ in Eq.~\eqref{eq:chic} actually does not contribute to the ground state energy. For the chiral spin superfluid, we find a spin-charge mixed Hamiltonian 
\bea 
{\cal H}_{\chi_s}  &=& \sum_{\nu = c,s}  \frac{1}{2} v_\nu \left[ K_\nu \Pi_\nu ^2 + \frac{1}{K_\nu} (\partial_x \theta_\nu)^2 \right] \nonumber \\ 
&& + \lambda \left[ \Pi_c \frac{\partial^3}{\partial x^3 } \theta_s + \Pi_s \frac{\partial^3}{\partial x^3 } \theta_c \right].
\eea
When compared to the chiral charge superfluid, the spin-charge mixing term leads to an additional energy-density correction 
\be 
\Delta {E}/L = - \frac{\lambda^2 \Lambda^6} {24 \pi}   \frac{ (K_c-K_s)^2}{ K_c K_s (v_c + v_s)}.
\label{eq:DeltaE} 
\ee  
and makes the chiral spin superfluid the actual ground state of our system. 

In optical lattice experiments, the chiral spin superfluid should be detectable using spin-resolved time-of-flight techniques as two spins spontaneously (quasi-) condense in different valleys, forming a momentum space  antiferromagnet. In the language of conformal field theory, the chiral spin superfluid is a critical phase with two gapless normal modes and is formally described by two Virasoro algebras with central charge $c=1$~\cite{1990_Frahm-Korepin-PRB}. 

\begin{figure}
\includegraphics[width=0.5\textwidth]{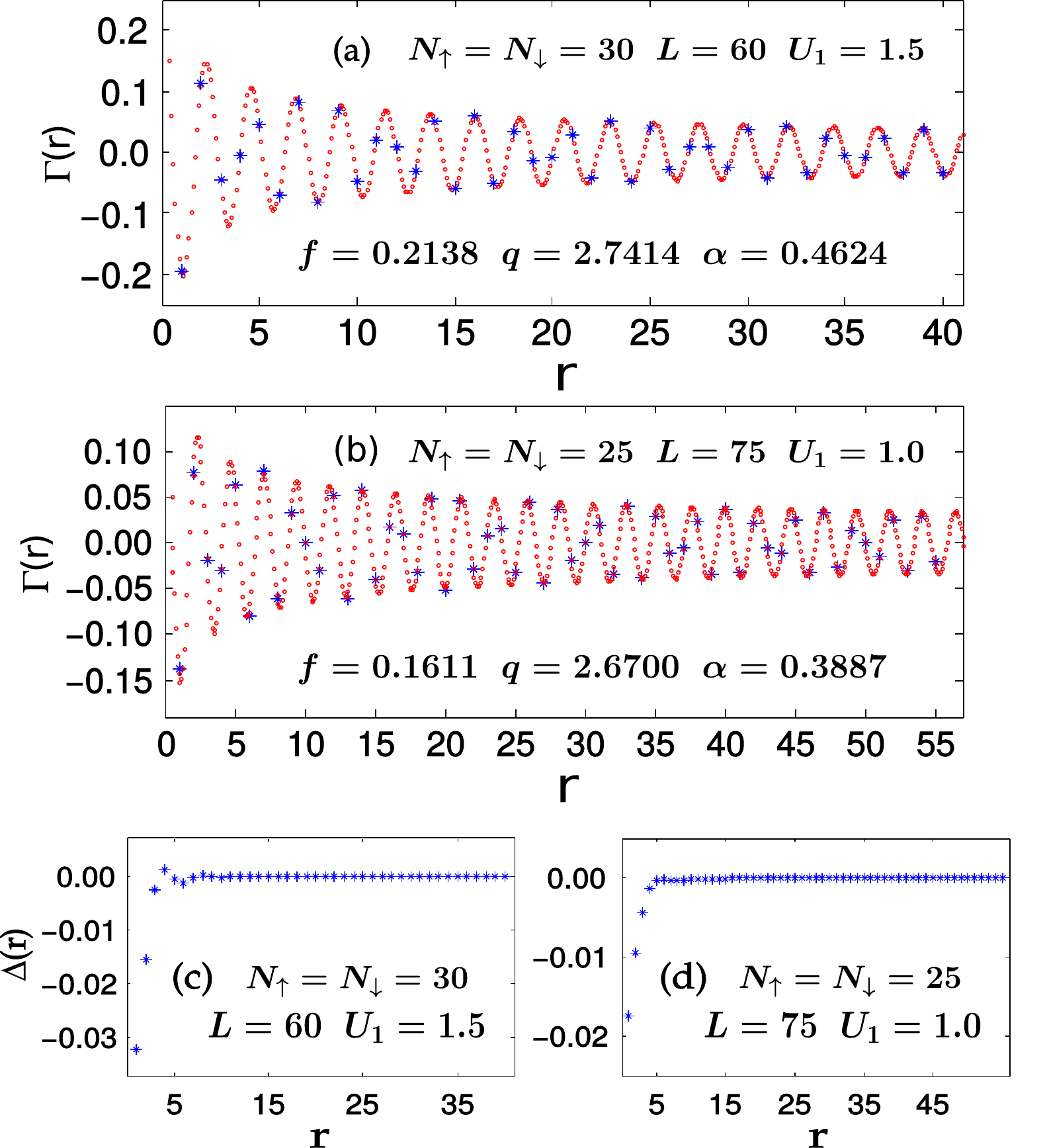}
\caption{The superfluid [(a) and (b)] and density [(c) and (d)] correlation functions for the $\pi$-flux triangular ladder. In the calculation of correlation function, we choose the left-most lattice site to be $m=10$ and the distance $r=n-m$. In panels (a) and (b), the blue stars are numerical results and the red dots are the least square fitting results using Eq.~\ref{MomentumFunction}. The fitting parameters are given in the panels.}
\label{Figure2}
\end{figure}

\medskip

\section{$\pi$-flux Triangular Ladder Model}

To further demonstrate the existence of the chiral spin superfluid, we choose a concrete optical lattice --- a $\pi$ flux triangular two-leg ladder (see the experiment in Ref.~\onlinecite{2013_Sengstock_NatPhys}), and carry out numerical calculations. As shown in Fig.~\ref{Figure1}, we consider a two-leg ladder with the legs labeled as ${\rm I}$ and ${\rm II}$, aligned along the $x$ direction. The number of sites along the $x$ direction is $L$ and these sites are indexed by $m$. The creation (annihilation) operators are denoted as $b^{\dagger}_{\sigma,\alpha,m}$ ($b_{\sigma,\alpha,m}$) with the spin, leg, and site indices, $\sigma=\uparrow,\downarrow$, $\alpha={\rm I},{\rm II}$, and $m\in[1,2,\cdots,L]$. The single-particle Hamiltonian consists of hopping terms connecting all the nearest neighbors with a {\em positive} coefficient as given by
\begin{eqnarray}
H_{0} &=& \sum_{\sigma=\uparrow,\downarrow} \sum_{m} \Big[ b^{\dagger}_{\sigma,{\rm I},m} b_{\sigma,{\rm I},m+1} + b^{\dagger}_{\sigma,{\rm I},m} b_{\sigma,{\rm II},m} \nonumber \\
      &+& b^{\dagger}_{\sigma,{\rm I},m} b_{\sigma,{\rm II},m+1} + b^{\dagger}_{\sigma,{\rm II},m} b_{\sigma,{\rm II},m+1} + {\rm H.c.} \Big]
\end{eqnarray}
The positive coefficient can be obtained using $\pi$ magnetic flux in each unit cell, which is impossible to reach in electronic systems but has been implemented in ultracold atoms~\cite{2013_Sengstock_NatPhys}. With periodic boundary condition along the $x$ direction, the single-particle Hamiltonian in momentum space reads 
\begin{eqnarray}
\left[
\begin{array}{cc}
2\cos{k} & 1+\cos{k}+i\sin{k} \\
1+\cos{k}-i\sin{k} & 2\cos{k} \\
\end{array}
\right]
\end{eqnarray}
The two Bloch bands have energy eigenvalues $E_{\pm}=2\cos{k}\pm\sqrt{\sin^2{k}+(1+\cos{k})^2}$. The lower band has two equal minimal values at momenta $Q_{\pm}=\pm\arccos(-7/8)\approx\pm{2.6362}$ rad and their degeneracy is protected by time-reversal symmetry. We study many-body systems with the numbers of bosons denoted as $N_{\sigma}$ and define the filling factor as $(N_{\uparrow}+N_{\downarrow})/(2L)$. The interactions between the bosons are described by the terms
\begin{eqnarray}
V &=& \sum_{\sigma=\uparrow,\downarrow} \sum_{\alpha={\rm I},{\rm II}} \sum_{m} U_{0} b^{\dagger}_{\sigma, \alpha,m} b^{\dagger}_{\sigma,\alpha,m} b_{\sigma,\alpha, m} b_{\sigma,\alpha,m} \nonumber \\
  &+& \sum_{\alpha={\rm I},{\rm II}} \sum_{m} U_{1} b^{\dagger}_{\uparrow,\alpha,m} b^{\dagger}_{\downarrow,\alpha,m} b_{\downarrow,\alpha,m} b_{\uparrow,\alpha,m} 
\end{eqnarray}
where $U_{0}=\infty$ (we choose  hard-core bosons for computational ease) and $U_{1}$ is a finite number. When computing the many-body ground state of our system using the DMRG method, we employ open boundary conditions as this is more efficient than periodic boundary conditions. We extract useful information only using the lattice sites far from the edges to suppress any edge effects.

\section{Numerical Results} 

\begin{figure*}
\includegraphics[width=0.9\textwidth]{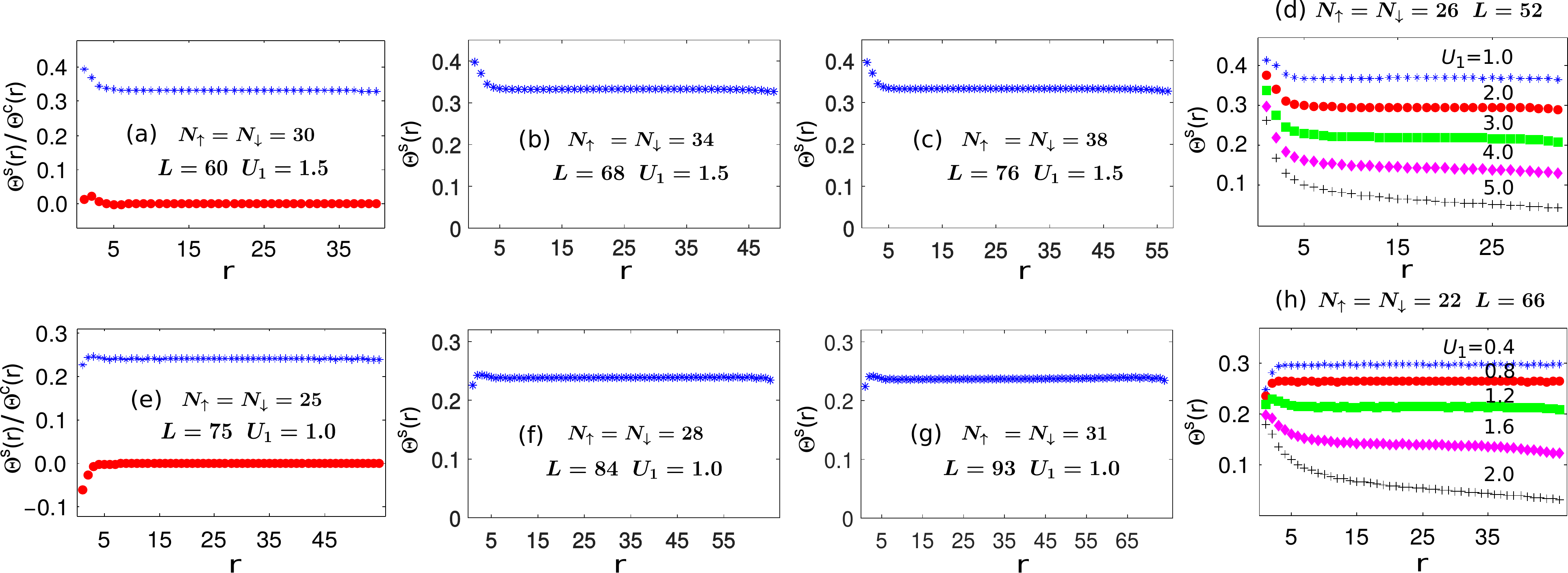}
\caption{The panels (a) and (b) show the charge (red dots) and spin (blue stars) current correlation functions for the same system as in Fig. \ref{Figure2}. The panels (c) and (d) show the spin current correlation functions for two systems at several different $U_{1}$ values. The left-most lattice site in calculating the correlation is chosen to be $m=10$ and the distance $r=n-m$.}
\label{Figure3}
\end{figure*}

We first analyze the physics of spinless bosons. If the bosons do not interact with each other, they would occupy one of the momenta $Q_{\pm}$ and we have a macroscopic ground state degeneracy. As the interaction is turned on, the bosons prefer to occupy one of the momenta $Q_{\pm}$ and produce a chiral condensate with spontaneous time-reversal symmetry breaking. For a two-component system with no inter-species interaction, the two components can condense in either of the two valleys so we have four degenerate ground states schematically denoted as $|(\uparrow, Q_{+}), (\downarrow, Q_{+})\rangle$, $|(\uparrow, Q_{+}), (\downarrow, Q_{-})\rangle$, $|(\uparrow, Q_{-}), (\downarrow, Q_{+})\rangle$, and $|(\uparrow, Q_{-}),(\downarrow, Q_{-})\rangle$, with $|(\sigma, Q), (\sigma', Q')\rangle$ referring to the spin component $\sigma$ ($\sigma'$) condensing at $Q$ ($Q'$). The degeneracy of $|(\uparrow, Q_{+}), (\downarrow, Q_{+})\rangle$ and $|(\uparrow, Q_{-}), (\downarrow, Q_{-})\rangle$ is protected by time-reversal symmetry, and the degeneracy of $|(\uparrow, Q_{+}), (\downarrow, Q_{-})\rangle$ and $|(\uparrow, Q_{-}), (\downarrow, Q_{+})\rangle$ is protected by inversion symmetry. The former pair of states posses chiral charge order while the latter pair of states posses chiral spin order.  These two orders are degenerate at the classical level, so the question to address is which order the inter-species interaction $U_1$ would select in the presence of strong quantum fluctuations.

We compute the ground states at $1/2$ and $1/3$ fillings (the physics at other fillings are qualitatively the same). In one dimension, a superfluid phase features a quasi-long-range order in the correlation $\Gamma_{mn}=\langle b^{\dagger}_{\sigma,\alpha,m} b_{\sigma,\alpha,n} \rangle$ due to strong quantum fluctuations. We show two examples of $\Gamma_{mn}$ for spin-up bosons on leg-I in Fig. \ref{Figure2} (a) and (b) and fit them according to 
\begin{eqnarray}
\Gamma_{mn}=f\cos[q(m-n)]/|m-n|^{\alpha}
\label{MomentumFunction}
\end{eqnarray}
In both cases, the coefficient $q$ is very close to $|Q_{\pm}|$ so we conclude that the ground states are indeed superfluids in which the bosons (quasi-) condense at $Q_{\pm}$. As another check, we have also computed the density-density correlation functions $\Delta_{mn}=\langle \rho_{\sigma,\alpha,m} \rho_{\sigma,\alpha,n} \rangle - \langle \rho_{\sigma,\alpha,m} \rangle \langle \rho_{\sigma,\alpha,n} \rangle $ ($\rho_{\sigma,\alpha,m}=b^{\dagger}_{\sigma,\alpha,m} b_{\sigma,\alpha,m}$). The two examples for spin-up bosons on leg-I shown in Fig. \ref{Figure2} (c) and (d) decay to zero very quickly so there is no long-range density order. Because the Hamiltonian is symmetric between spin-up and spin-down bosons, the same correlation functions are expected for spin-down bosons, which we have confirmed explicitly.

\begin{figure}
\includegraphics[width=0.5\textwidth]{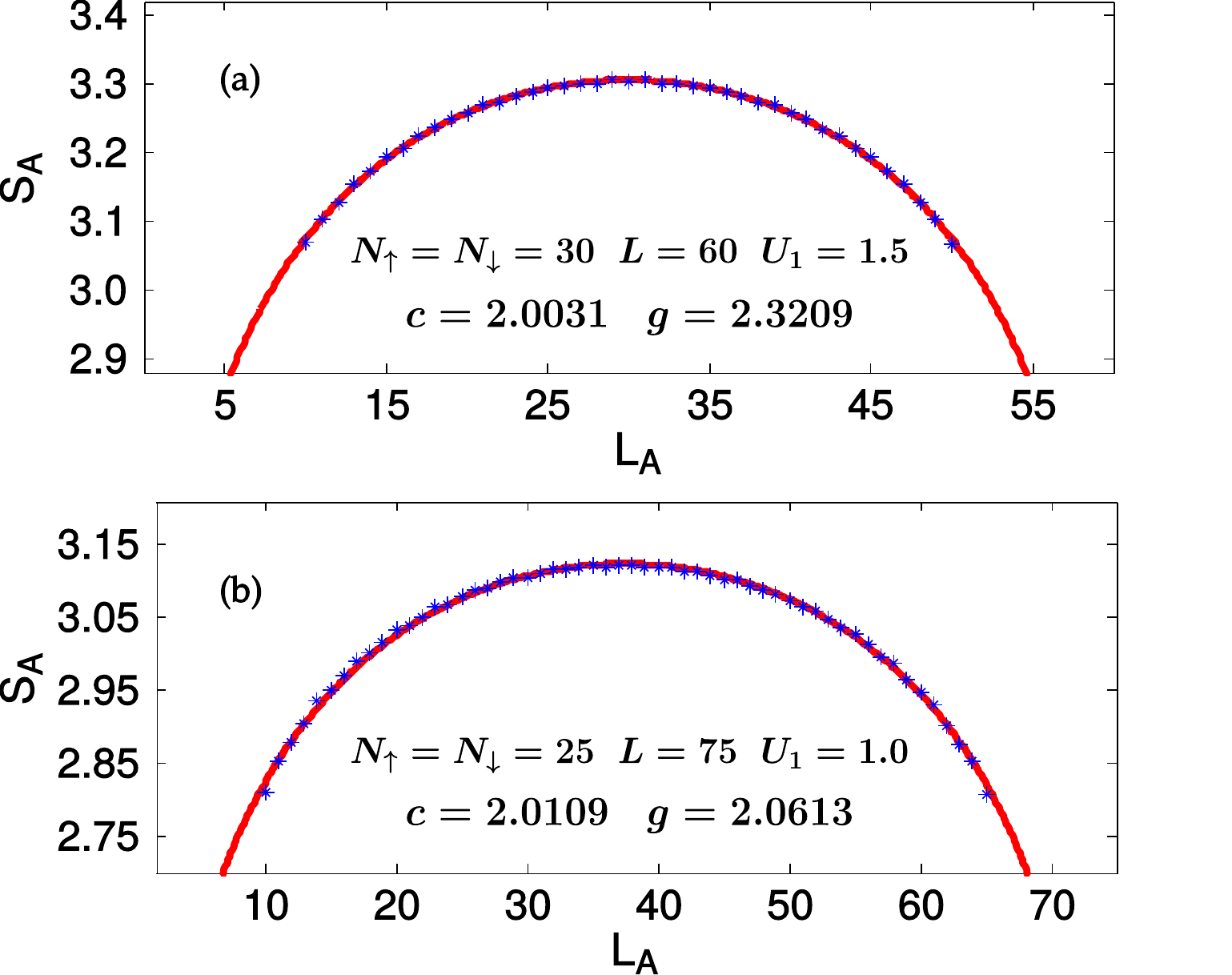}
\caption{The von Neumann entanglement entropy. The blue stars are numerical results and the red lines are the least square fitting results using Eq. \ref{EntropyFunction} without the oscillating term. The fitting parameters are given in the panels.}
\label{Figure4}
\end{figure}

The superfluid and density correlation functions demonstrate that the ground states are superfluid but they do not tell us whether the two components condense in the same or different valleys. To distinguish between chiral charge and chiral spin orders, we define the current operator on the bond $m$ as $J_{{\sigma}m} = i( b^{\dagger}_{\sigma,{\rm I},m} b_{\sigma,{\rm II},m} - b^{\dagger}_{\sigma,{\rm II},m} b_{\sigma,{\rm I},m} )$ and compute its correlation functions. The two types of chiral orders will give rise to two different long-range ordered correlation functions---$\Theta^{c}_{mn}=\langle J^{c}_{m} J^{c}_{n} \rangle$ and $\Theta^{s}_{mn}=\langle J^{s}_{m} J^{s}_{n} \rangle$, respectively, where $J^{c}_{m}=J_{{\uparrow}m}+J_{{\downarrow}m}$ is the charge current operator and $J^{s}_{m}=J_{{\uparrow}m}-J_{{\downarrow}m}$ is the spin current operator. The chiral spin ordered phase breaks the inversion symmetry spontaneously and the spin current correlation function exhibits true long-range order because the broken symmetry is  discrete. Fig. \ref{Figure3}  shows the correlation functions $\Theta^{c}_{mn}$ and $\Theta^{s}_{mn}$. We find that $\Theta^{c}_{mn}$ decays to zero quickly as $m-n$ increases while $\Theta^{s}_{mn}$ saturates to a constant value in the bulk. The quantity $\Theta^{s}_{mn}$ is an ``order parameter" to identify the regime where the chiral spin condensate is stabilized. Fig. \ref{Figure3} (c) and (d) show $\Theta^{s}_{mn}$ for different  interaction strength $U_{1}$. In both cases, we find that as we increase $U_1$ the ground state exhibits chiral spin current when $U_1$  is not too large, but eventually $\Theta^{s}_{mn}$ becomes a decaying function of $m-n$ (meanwhile $\Theta^{c}_{mn}$ always decay to zero quickly). This implies that the chiral spin order gets weaker as the interspecies interaction strength increases, which is consistent with the field theory results in Eq.~\eqref{eq:DeltaE}. The large $U_1$ regime in our system is expected to resemble similar physics as described in Ref. \onlinecite{2004_Kuklov_PRL}.

To confirm that the low-energy effective field theory is indeed a two-component Luttinger liquid, we calculate the scaling of the entanglement entropy using DMRG. We choose a subsystem $A$ with $L_{A}$ sites on the left of the system, trace out the other sites to obtain the reduced density matrix $\rho_{A}$ of $A$, and compute the von Neumann entropy $S=-{\rm Tr}(\rho_{A} \ln\rho_{A})$. For a system that is open in the $x$ direction, $S$ is predicted to take a functional form of~\cite{2004_Calabrese-Cardy}
\begin{eqnarray}
S(L_{A}) = \frac{c}{6} \ln\left[ \frac{L}{\pi} \sin\left(\pi\frac{L_{A}}{L}\right) \right] + g + F
\label{EntropyFunction}
\end{eqnarray}
based on conformal field theory, where $g$ is a constant and $F$ is a non-universal oscillating term. In our model, we find  that the $F$ term becomes less important as the system size increases so the value of $c$ can be extracted using sufficiently large systems. To avoid edge effects, we discard the data points for which $L_{A}$ is close to zero or $L$ and only use those in the middle. The two examples shown in Fig. \ref{Figure4} both give central charge $2$  as expected for a two-component Luttinger liquid.

\section{Conclusion}

Based on effective field theory analysis and numerical simulations, we have established that the ground state of two-component bosons in a one-dimensional lattice with double-valley band could be a chiral spin superfluid. This phase spontaneously breaks inversion symmetry and exhibits chiral spin loop currents. Our predictions should be readily testable in ultracold atoms using spin-resolved time-of-flight techniques since the system exhibits momentum space antiferromagnetism. The chiral spin condensate may lead to applications in quantum information processing and topological state engineering. For example, it may be used as building blocks of chiral spin networks for multiparticle entanglement generation~\cite{2015_Pichler_Zoller_PRA}. In Bose-Fermi mixtures, chiral spin condensate of bosons may provide a background for the fermions to form topological phases. The prospect of inducing topological phases from spontaneous symmetry breaking ~\cite{2008_Raghu_Qi_PRL,2009_Sun_PRL,2015_Li_NatComm,2015_Xu_Li_PRL,2016_Zhu_Sheng_PRB} is worth further exploration.

The $\pi$-flux triangular ladder model may also host other interesting physics. With optical lattices~\cite{2013_Sengstock_NatPhys}, the hopping constants along the legs and between the legs could in principle be changed independently. The interaction strengths are also tunable by changing lattice depth and by Feshbach resonances. Varying these parameters, a even richer phase diagram is anticipated. In particular, the bosons may form a Mott insulator rather than a superfluid at integer filling factors. It remains to be seen if the insulating state can also possess nontrivial chiral spin order.

\section{Acknowledgement}

YHW thanks Hong-Hao Tu for helpful discussions. Work at MPQ was supproted by the European Union Project {\em Simulators and Interfaces with Quantum Systems}. Work at Maryland was supported by JQI-NSF-PFC,  LPS-MPO-CMTC, and the PFC seed grant ``Emergent phonemena in interacting spin-orbit coupled gases" (XL). XL acknowledges KITP for hospitality where part of the manuscript was finished. XL is supported by  National Program on Key Basic Research Project of China (Grant No. 2017YFA0304204), National Natural Science Foundation of China  (Grant No. 117740067),  the Thousand-Youth-Talent Program of China. 

\bibliographystyle{apsrev4-1}
\bibliography{references}

\begin{thebibliography}{45}%
\makeatletter
\providecommand \@ifxundefined [1]{%
 \@ifx{#1\undefined}
}%
\providecommand \@ifnum [1]{%
 \ifnum #1\expandafter \@firstoftwo
 \else \expandafter \@secondoftwo
 \fi
}%
\providecommand \@ifx [1]{%
 \ifx #1\expandafter \@firstoftwo
 \else \expandafter \@secondoftwo
 \fi
}%
\providecommand \natexlab [1]{#1}%
\providecommand \enquote  [1]{``#1''}%
\providecommand \bibnamefont  [1]{#1}%
\providecommand \bibfnamefont [1]{#1}%
\providecommand \citenamefont [1]{#1}%
\providecommand \href@noop [0]{\@secondoftwo}%
\providecommand \href [0]{\begingroup \@sanitize@url \@href}%
\providecommand \@href[1]{\@@startlink{#1}\@@href}%
\providecommand \@@href[1]{\endgroup#1\@@endlink}%
\providecommand \@sanitize@url [0]{\catcode `\\12\catcode `\$12\catcode
  `\&12\catcode `\#12\catcode `\^12\catcode `\_12\catcode `\%12\relax}%
\providecommand \@@startlink[1]{}%
\providecommand \@@endlink[0]{}%
\providecommand \url  [0]{\begingroup\@sanitize@url \@url }%
\providecommand \@url [1]{\endgroup\@href {#1}{\urlprefix }}%
\providecommand \urlprefix  [0]{URL }%
\providecommand \Eprint [0]{\href }%
\providecommand \doibase [0]{http://dx.doi.org/}%
\providecommand \selectlanguage [0]{\@gobble}%
\providecommand \bibinfo  [0]{\@secondoftwo}%
\providecommand \bibfield  [0]{\@secondoftwo}%
\providecommand \translation [1]{[#1]}%
\providecommand \BibitemOpen [0]{}%
\providecommand \bibitemStop [0]{}%
\providecommand \bibitemNoStop [0]{.\EOS\space}%
\providecommand \EOS [0]{\spacefactor3000\relax}%
\providecommand \BibitemShut  [1]{\csname bibitem#1\endcsname}%
\let\auto@bib@innerbib\@empty
\bibitem [{\citenamefont {Jaksch}\ \emph {et~al.}(1998)\citenamefont {Jaksch},
  \citenamefont {Bruder}, \citenamefont {Cirac}, \citenamefont {Gardiner},\
  and\ \citenamefont {Zoller}}]{1998_Zoller_Jaksch_PRL}%
  \BibitemOpen
  \bibfield  {author} {\bibinfo {author} {\bibfnamefont {D.}~\bibnamefont
  {Jaksch}}, \bibinfo {author} {\bibfnamefont {C.}~\bibnamefont {Bruder}},
  \bibinfo {author} {\bibfnamefont {J.~I.}\ \bibnamefont {Cirac}}, \bibinfo
  {author} {\bibfnamefont {C.~W.}\ \bibnamefont {Gardiner}}, \ and\ \bibinfo
  {author} {\bibfnamefont {P.}~\bibnamefont {Zoller}},\ }\href {\doibase
  10.1103/PhysRevLett.81.3108} {\bibfield  {journal} {\bibinfo  {journal}
  {Phys. Rev. Lett.}\ }\textbf {\bibinfo {volume} {81}},\ \bibinfo {pages}
  {3108} (\bibinfo {year} {1998})}\BibitemShut {NoStop}%
\bibitem [{\citenamefont {Hofstetter}\ \emph {et~al.}(2002)\citenamefont
  {Hofstetter}, \citenamefont {Cirac}, \citenamefont {Zoller}, \citenamefont
  {Demler},\ and\ \citenamefont {Lukin}}]{2002_Hofstetter_Cirac_PRL}%
  \BibitemOpen
  \bibfield  {author} {\bibinfo {author} {\bibfnamefont {W.}~\bibnamefont
  {Hofstetter}}, \bibinfo {author} {\bibfnamefont {J.~I.}\ \bibnamefont
  {Cirac}}, \bibinfo {author} {\bibfnamefont {P.}~\bibnamefont {Zoller}},
  \bibinfo {author} {\bibfnamefont {E.}~\bibnamefont {Demler}}, \ and\ \bibinfo
  {author} {\bibfnamefont {M.~D.}\ \bibnamefont {Lukin}},\ }\href {\doibase
  10.1103/PhysRevLett.89.220407} {\bibfield  {journal} {\bibinfo  {journal}
  {Phys. Rev. Lett.}\ }\textbf {\bibinfo {volume} {89}},\ \bibinfo {pages}
  {220407} (\bibinfo {year} {2002})}\BibitemShut {NoStop}%
\bibitem [{\citenamefont {Lewenstein}\ \emph {et~al.}(2007)\citenamefont
  {Lewenstein}, \citenamefont {Sanpera}, \citenamefont {Ahufinger},
  \citenamefont {Damski}, \citenamefont {Sen(De)},\ and\ \citenamefont
  {Sen}}]{2007_Lewenstein_AP}%
  \BibitemOpen
  \bibfield  {author} {\bibinfo {author} {\bibfnamefont {M.}~\bibnamefont
  {Lewenstein}}, \bibinfo {author} {\bibfnamefont {A.}~\bibnamefont {Sanpera}},
  \bibinfo {author} {\bibfnamefont {V.}~\bibnamefont {Ahufinger}}, \bibinfo
  {author} {\bibfnamefont {B.}~\bibnamefont {Damski}}, \bibinfo {author}
  {\bibfnamefont {A.}~\bibnamefont {Sen(De)}}, \ and\ \bibinfo {author}
  {\bibfnamefont {U.}~\bibnamefont {Sen}},\ }\href {\doibase
  10.1080/00018730701223200} {\bibfield  {journal} {\bibinfo  {journal}
  {Advances in Physics}\ }\textbf {\bibinfo {volume} {56}},\ \bibinfo {pages}
  {243} (\bibinfo {year} {2007})}\BibitemShut {NoStop}%
\bibitem [{\citenamefont {Bloch}\ \emph {et~al.}(2008)\citenamefont {Bloch},
  \citenamefont {Dalibard},\ and\ \citenamefont
  {Zwerger}}]{2008_Bloch_Dalibard_RMP}%
  \BibitemOpen
  \bibfield  {author} {\bibinfo {author} {\bibfnamefont {I.}~\bibnamefont
  {Bloch}}, \bibinfo {author} {\bibfnamefont {J.}~\bibnamefont {Dalibard}}, \
  and\ \bibinfo {author} {\bibfnamefont {W.}~\bibnamefont {Zwerger}},\ }\href
  {\doibase 10.1103/RevModPhys.80.885} {\bibfield  {journal} {\bibinfo
  {journal} {Rev. Mod. Phys.}\ }\textbf {\bibinfo {volume} {80}},\ \bibinfo
  {pages} {885} (\bibinfo {year} {2008})}\BibitemShut {NoStop}%
\bibitem [{\citenamefont {Esslinger}(2010)}]{2010_Esslinger_CMP}%
  \BibitemOpen
  \bibfield  {author} {\bibinfo {author} {\bibfnamefont {T.}~\bibnamefont
  {Esslinger}},\ }\href
  {http://www.annualreviews.org/doi/10.1146/annurev-conmatphys-070909-104059}
  {\bibfield  {journal} {\bibinfo  {journal} {Ann. Rev. Cond. Matt. Phys.}\
  }\textbf {\bibinfo {volume} {1}},\ \bibinfo {pages} {129} (\bibinfo {year}
  {2010})}\BibitemShut {NoStop}%
\bibitem [{\citenamefont {Dutta}\ \emph {et~al.}(2015)\citenamefont {Dutta},
  \citenamefont {Gajda}, \citenamefont {Hauke}, \citenamefont {Lewenstein},
  \citenamefont {Lühmann}, \citenamefont {Malomed}, \citenamefont
  {Sowiński},\ and\ \citenamefont {Zakrzewski}}]{2015_Lewenstein_RPP}%
  \BibitemOpen
  \bibfield  {author} {\bibinfo {author} {\bibfnamefont {O.}~\bibnamefont
  {Dutta}}, \bibinfo {author} {\bibfnamefont {M.}~\bibnamefont {Gajda}},
  \bibinfo {author} {\bibfnamefont {P.}~\bibnamefont {Hauke}}, \bibinfo
  {author} {\bibfnamefont {M.}~\bibnamefont {Lewenstein}}, \bibinfo {author}
  {\bibfnamefont {D.-S.}\ \bibnamefont {Lühmann}}, \bibinfo {author}
  {\bibfnamefont {B.~A.}\ \bibnamefont {Malomed}}, \bibinfo {author}
  {\bibfnamefont {T.}~\bibnamefont {Sowiński}}, \ and\ \bibinfo {author}
  {\bibfnamefont {J.}~\bibnamefont {Zakrzewski}},\ }\href
  {http://stacks.iop.org/0034-4885/78/i=6/a=066001} {\bibfield  {journal}
  {\bibinfo  {journal} {Rep. Prog. Phys.}\ }\textbf {\bibinfo
  {volume} {78}},\ \bibinfo {pages} {066001} (\bibinfo {year}
  {2015})}\BibitemShut {NoStop}%
\bibitem [{\citenamefont {Li}\ and\ \citenamefont
  {Liu}(2016)}]{2016_Li_Liu_RPP}%
  \BibitemOpen
  \bibfield  {author} {\bibinfo {author} {\bibfnamefont {X.}~\bibnamefont
  {Li}}\ and\ \bibinfo {author} {\bibfnamefont {W.~V.}\ \bibnamefont {Liu}},\
  }\href {http://stacks.iop.org/0034-4885/79/i=11/a=116401} {\bibfield
  {journal} {\bibinfo  {journal} {Rep. Prog. Phys.}\ }\textbf
  {\bibinfo {volume} {79}},\ \bibinfo {pages} {116401} (\bibinfo {year}
  {2016})}\BibitemShut {NoStop}%
\bibitem [{\citenamefont {Dalibard}\ \emph {et~al.}(2011)\citenamefont
  {Dalibard}, \citenamefont {Gerbier}, \citenamefont
  {Juzeli\ifmmode~\bar{u}\else \={u}\fi{}nas},\ and\ \citenamefont
  {\"Ohberg}}]{2011_Dalibard_Gerbier_RMP}%
  \BibitemOpen
  \bibfield  {author} {\bibinfo {author} {\bibfnamefont {J.}~\bibnamefont
  {Dalibard}}, \bibinfo {author} {\bibfnamefont {F.}~\bibnamefont {Gerbier}},
  \bibinfo {author} {\bibfnamefont {G.}~\bibnamefont
  {Juzeli\ifmmode~\bar{u}\else \={u}\fi{}nas}}, \ and\ \bibinfo {author}
  {\bibfnamefont {P.}~\bibnamefont {\"Ohberg}},\ }\href {\doibase
  10.1103/RevModPhys.83.1523} {\bibfield  {journal} {\bibinfo  {journal} {Rev.
  Mod. Phys.}\ }\textbf {\bibinfo {volume} {83}},\ \bibinfo {pages} {1523}
  (\bibinfo {year} {2011})}\BibitemShut {NoStop}%
\bibitem [{\citenamefont {Parker}\ \emph {et~al.}(2013)\citenamefont {Parker},
  \citenamefont {Ha},\ and\ \citenamefont {Chin}}]{2013_Chin_NatPhys}%
  \BibitemOpen
  \bibfield  {author} {\bibinfo {author} {\bibfnamefont {C.~V.}\ \bibnamefont
  {Parker}}, \bibinfo {author} {\bibfnamefont {L.-C.}\ \bibnamefont {Ha}}, \
  and\ \bibinfo {author} {\bibfnamefont {C.}~\bibnamefont {Chin}},\ }\href
  {http://dx.doi.org/10.1038/nphys2789} {\bibfield  {journal} {\bibinfo
  {journal} {Nat Phys}\ }\textbf {\bibinfo {volume} {9}},\ \bibinfo {pages}
  {769} (\bibinfo {year} {2013})}\BibitemShut {NoStop}%
\bibitem [{\citenamefont {Struck}\ \emph {et~al.}(2013)\citenamefont {Struck},
  \citenamefont {Weinberg}, \citenamefont {Olschlager}, \citenamefont
  {Windpassinger}, \citenamefont {Simonet}, \citenamefont {Sengstock},
  \citenamefont {Hoppner}, \citenamefont {Hauke}, \citenamefont {Eckardt},
  \citenamefont {Lewenstein},\ and\ \citenamefont
  {Mathey}}]{2013_Sengstock_NatPhys}%
  \BibitemOpen
  \bibfield  {author} {\bibinfo {author} {\bibfnamefont {J.}~\bibnamefont
  {Struck}}, \bibinfo {author} {\bibfnamefont {M.}~\bibnamefont {Weinberg}},
  \bibinfo {author} {\bibfnamefont {C.}~\bibnamefont {Olschlager}}, \bibinfo
  {author} {\bibfnamefont {P.}~\bibnamefont {Windpassinger}}, \bibinfo {author}
  {\bibfnamefont {J.}~\bibnamefont {Simonet}}, \bibinfo {author} {\bibfnamefont
  {K.}~\bibnamefont {Sengstock}}, \bibinfo {author} {\bibfnamefont
  {R.}~\bibnamefont {Hoppner}}, \bibinfo {author} {\bibfnamefont
  {P.}~\bibnamefont {Hauke}}, \bibinfo {author} {\bibfnamefont
  {A.}~\bibnamefont {Eckardt}}, \bibinfo {author} {\bibfnamefont
  {M.}~\bibnamefont {Lewenstein}}, \ and\ \bibinfo {author} {\bibfnamefont
  {L.}~\bibnamefont {Mathey}},\ }\href {http://dx.doi.org/10.1038/nphys2750}
  {\bibfield  {journal} {\bibinfo  {journal} {Nat Phys}\ }\textbf {\bibinfo
  {volume} {9}},\ \bibinfo {pages} {738} (\bibinfo {year} {2013})}\BibitemShut
  {NoStop}%
\bibitem [{\citenamefont {Mishra}\ \emph {et~al.}(2013)\citenamefont {Mishra},
  \citenamefont {Pai}, \citenamefont {Mukerjee},\ and\ \citenamefont
  {Paramekanti}}]{2013_Paramekanti_PRB}%
  \BibitemOpen
  \bibfield  {author} {\bibinfo {author} {\bibfnamefont {T.}~\bibnamefont
  {Mishra}}, \bibinfo {author} {\bibfnamefont {R.~V.}\ \bibnamefont {Pai}},
  \bibinfo {author} {\bibfnamefont {S.}~\bibnamefont {Mukerjee}}, \ and\
  \bibinfo {author} {\bibfnamefont {A.}~\bibnamefont {Paramekanti}},\ }\href
  {\doibase 10.1103/PhysRevB.87.174504} {\bibfield  {journal} {\bibinfo
  {journal} {Phys. Rev. B}\ }\textbf {\bibinfo {volume} {87}},\ \bibinfo
  {pages} {174504} (\bibinfo {year} {2013})}\BibitemShut {NoStop}%
\bibitem [{\citenamefont {{Edmonds}}\ \emph {et~al.}(2015)\citenamefont
  {{Edmonds}}, \citenamefont {{Valiente}},\ and\ \citenamefont
  {{{\"O}hberg}}}]{2015_Edmonds_EPL}%
  \BibitemOpen
  \bibfield  {author} {\bibinfo {author} {\bibfnamefont {M.~J.}\ \bibnamefont
  {{Edmonds}}}, \bibinfo {author} {\bibfnamefont {M.}~\bibnamefont
  {{Valiente}}}, \ and\ \bibinfo {author} {\bibfnamefont {P.}~\bibnamefont
  {{{\"O}hberg}}},\ }\href {\doibase 10.1209/0295-5075/110/36004} {\bibfield
  {journal} {\bibinfo  {journal} {Europhysics Letters}\ }\textbf {\bibinfo
  {volume} {110}},\ \bibinfo {eid} {36004} (\bibinfo {year}
  {2015})}\BibitemShut {NoStop}%
\bibitem [{\citenamefont {Zaletel}\ \emph {et~al.}(2014)\citenamefont
  {Zaletel}, \citenamefont {Parameswaran}, \citenamefont {R\"uegg},\ and\
  \citenamefont {Altman}}]{2014_Zaletel_PRB}%
  \BibitemOpen
  \bibfield  {author} {\bibinfo {author} {\bibfnamefont {M.~P.}\ \bibnamefont
  {Zaletel}}, \bibinfo {author} {\bibfnamefont {S.~A.}\ \bibnamefont
  {Parameswaran}}, \bibinfo {author} {\bibfnamefont {A.}~\bibnamefont
  {R\"uegg}}, \ and\ \bibinfo {author} {\bibfnamefont {E.}~\bibnamefont
  {Altman}},\ }\href {\doibase 10.1103/PhysRevB.89.155142} {\bibfield
  {journal} {\bibinfo  {journal} {Phys. Rev. B}\ }\textbf {\bibinfo {volume}
  {89}},\ \bibinfo {pages} {155142} (\bibinfo {year} {2014})}\BibitemShut
  {NoStop}%
\bibitem [{\citenamefont {Klitzing}\ \emph {et~al.}(1980)\citenamefont
  {Klitzing}, \citenamefont {Dorda},\ and\ \citenamefont
  {Pepper}}]{1980_Klitzing_PRL}%
  \BibitemOpen
  \bibfield  {author} {\bibinfo {author} {\bibfnamefont {K.~v.}\ \bibnamefont
  {Klitzing}}, \bibinfo {author} {\bibfnamefont {G.}~\bibnamefont {Dorda}}, \
  and\ \bibinfo {author} {\bibfnamefont {M.}~\bibnamefont {Pepper}},\ }\href
  {\doibase 10.1103/PhysRevLett.45.494} {\bibfield  {journal} {\bibinfo
  {journal} {Phys. Rev. Lett.}\ }\textbf {\bibinfo {volume} {45}},\ \bibinfo
  {pages} {494} (\bibinfo {year} {1980})}\BibitemShut {NoStop}%
\bibitem [{\citenamefont {Ruseckas}\ \emph {et~al.}(2005)\citenamefont
  {Ruseckas}, \citenamefont {Juzeli\ifmmode~\bar{u}\else \={u}\fi{}nas},
  \citenamefont {\"Ohberg},\ and\ \citenamefont
  {Fleischhauer}}]{2005_Ruseckas_PRL}%
  \BibitemOpen
  \bibfield  {author} {\bibinfo {author} {\bibfnamefont {J.}~\bibnamefont
  {Ruseckas}}, \bibinfo {author} {\bibfnamefont {G.}~\bibnamefont
  {Juzeli\ifmmode~\bar{u}\else \={u}\fi{}nas}}, \bibinfo {author}
  {\bibfnamefont {P.}~\bibnamefont {\"Ohberg}}, \ and\ \bibinfo {author}
  {\bibfnamefont {M.}~\bibnamefont {Fleischhauer}},\ }\href {\doibase
  10.1103/PhysRevLett.95.010404} {\bibfield  {journal} {\bibinfo  {journal}
  {Phys. Rev. Lett.}\ }\textbf {\bibinfo {volume} {95}},\ \bibinfo {pages}
  {010404} (\bibinfo {year} {2005})}\BibitemShut {NoStop}%
\bibitem [{\citenamefont {Osterloh}\ \emph {et~al.}(2005)\citenamefont
  {Osterloh}, \citenamefont {Baig}, \citenamefont {Santos}, \citenamefont
  {Zoller},\ and\ \citenamefont {Lewenstein}}]{2005_Osterloh_Zoller_PRL}%
  \BibitemOpen
  \bibfield  {author} {\bibinfo {author} {\bibfnamefont {K.}~\bibnamefont
  {Osterloh}}, \bibinfo {author} {\bibfnamefont {M.}~\bibnamefont {Baig}},
  \bibinfo {author} {\bibfnamefont {L.}~\bibnamefont {Santos}}, \bibinfo
  {author} {\bibfnamefont {P.}~\bibnamefont {Zoller}}, \ and\ \bibinfo {author}
  {\bibfnamefont {M.}~\bibnamefont {Lewenstein}},\ }\href {\doibase
  10.1103/PhysRevLett.95.010403} {\bibfield  {journal} {\bibinfo  {journal}
  {Phys. Rev. Lett.}\ }\textbf {\bibinfo {volume} {95}},\ \bibinfo {pages}
  {010403} (\bibinfo {year} {2005})}\BibitemShut {NoStop}%
\bibitem [{\citenamefont {Liu}\ \emph {et~al.}(2009)\citenamefont {Liu},
  \citenamefont {Borunda}, \citenamefont {Liu},\ and\ \citenamefont
  {Sinova}}]{2009_Liu_PRL}%
  \BibitemOpen
  \bibfield  {author} {\bibinfo {author} {\bibfnamefont {X.-J.}\ \bibnamefont
  {Liu}}, \bibinfo {author} {\bibfnamefont {M.~F.}\ \bibnamefont {Borunda}},
  \bibinfo {author} {\bibfnamefont {X.}~\bibnamefont {Liu}}, \ and\ \bibinfo
  {author} {\bibfnamefont {J.}~\bibnamefont {Sinova}},\ }\href {\doibase
  10.1103/PhysRevLett.102.046402} {\bibfield  {journal} {\bibinfo  {journal}
  {Phys. Rev. Lett.}\ }\textbf {\bibinfo {volume} {102}},\ \bibinfo {pages}
  {046402} (\bibinfo {year} {2009})}\BibitemShut {NoStop}%
\bibitem [{\citenamefont {Lin}\ \emph {et~al.}(2011)\citenamefont {Lin},
  \citenamefont {Jimenez-Garcia},\ and\ \citenamefont
  {Spielman}}]{2011_Lin_Spielman_Nature}%
  \BibitemOpen
  \bibfield  {author} {\bibinfo {author} {\bibfnamefont {Y.-J.}\ \bibnamefont
  {Lin}}, \bibinfo {author} {\bibfnamefont {K.}~\bibnamefont {Jimenez-Garcia}},
  \ and\ \bibinfo {author} {\bibfnamefont {I.~B.}\ \bibnamefont {Spielman}},\
  }\href {http://dx.doi.org/10.1038/nature09887} {\bibfield  {journal}
  {\bibinfo  {journal} {Nature}\ }\textbf {\bibinfo {volume} {471}},\ \bibinfo
  {pages} {83} (\bibinfo {year} {2011})}\BibitemShut {NoStop}%
\bibitem [{\citenamefont {Zhang}\ \emph {et~al.}(2012)\citenamefont {Zhang},
  \citenamefont {Ji}, \citenamefont {Chen}, \citenamefont {Zhang},
  \citenamefont {Du}, \citenamefont {Yan}, \citenamefont {Pan}, \citenamefont
  {Zhao}, \citenamefont {Deng}, \citenamefont {Zhai}, \citenamefont {Chen},\
  and\ \citenamefont {Pan}}]{2012_Pan_SOC}%
  \BibitemOpen
  \bibfield  {author} {\bibinfo {author} {\bibfnamefont {J.-Y.}\ \bibnamefont
  {Zhang}}, \bibinfo {author} {\bibfnamefont {S.-C.}\ \bibnamefont {Ji}},
  \bibinfo {author} {\bibfnamefont {Z.}~\bibnamefont {Chen}}, \bibinfo {author}
  {\bibfnamefont {L.}~\bibnamefont {Zhang}}, \bibinfo {author} {\bibfnamefont
  {Z.-D.}\ \bibnamefont {Du}}, \bibinfo {author} {\bibfnamefont
  {B.}~\bibnamefont {Yan}}, \bibinfo {author} {\bibfnamefont {G.-S.}\
  \bibnamefont {Pan}}, \bibinfo {author} {\bibfnamefont {B.}~\bibnamefont
  {Zhao}}, \bibinfo {author} {\bibfnamefont {Y.-J.}\ \bibnamefont {Deng}},
  \bibinfo {author} {\bibfnamefont {H.}~\bibnamefont {Zhai}}, \bibinfo {author}
  {\bibfnamefont {S.}~\bibnamefont {Chen}}, \ and\ \bibinfo {author}
  {\bibfnamefont {J.-W.}\ \bibnamefont {Pan}},\ }\href {\doibase
  10.1103/PhysRevLett.109.115301} {\bibfield  {journal} {\bibinfo  {journal}
  {Phys. Rev. Lett.}\ }\textbf {\bibinfo {volume} {109}},\ \bibinfo {pages}
  {115301} (\bibinfo {year} {2012})}\BibitemShut {NoStop}%
\bibitem [{\citenamefont {Wang}\ \emph {et~al.}(2012)\citenamefont {Wang},
  \citenamefont {Yu}, \citenamefont {Fu}, \citenamefont {Miao}, \citenamefont
  {Huang}, \citenamefont {Chai}, \citenamefont {Zhai},\ and\ \citenamefont
  {Zhang}}]{2012_Zhang_Zhai_PRL}%
  \BibitemOpen
  \bibfield  {author} {\bibinfo {author} {\bibfnamefont {P.}~\bibnamefont
  {Wang}}, \bibinfo {author} {\bibfnamefont {Z.-Q.}\ \bibnamefont {Yu}},
  \bibinfo {author} {\bibfnamefont {Z.}~\bibnamefont {Fu}}, \bibinfo {author}
  {\bibfnamefont {J.}~\bibnamefont {Miao}}, \bibinfo {author} {\bibfnamefont
  {L.}~\bibnamefont {Huang}}, \bibinfo {author} {\bibfnamefont
  {S.}~\bibnamefont {Chai}}, \bibinfo {author} {\bibfnamefont {H.}~\bibnamefont
  {Zhai}}, \ and\ \bibinfo {author} {\bibfnamefont {J.}~\bibnamefont {Zhang}},\
  }\href {\doibase 10.1103/PhysRevLett.109.095301} {\bibfield  {journal}
  {\bibinfo  {journal} {Phys. Rev. Lett.}\ }\textbf {\bibinfo {volume} {109}},\
  \bibinfo {pages} {095301} (\bibinfo {year} {2012})}\BibitemShut {NoStop}%
\bibitem [{\citenamefont {Cheuk}\ \emph {et~al.}(2012)\citenamefont {Cheuk},
  \citenamefont {Sommer}, \citenamefont {Hadzibabic}, \citenamefont {Yefsah},
  \citenamefont {Bakr},\ and\ \citenamefont
  {Zwierlein}}]{2012_Cheuk_Zwierlein_PRL}%
  \BibitemOpen
  \bibfield  {author} {\bibinfo {author} {\bibfnamefont {L.~W.}\ \bibnamefont
  {Cheuk}}, \bibinfo {author} {\bibfnamefont {A.~T.}\ \bibnamefont {Sommer}},
  \bibinfo {author} {\bibfnamefont {Z.}~\bibnamefont {Hadzibabic}}, \bibinfo
  {author} {\bibfnamefont {T.}~\bibnamefont {Yefsah}}, \bibinfo {author}
  {\bibfnamefont {W.~S.}\ \bibnamefont {Bakr}}, \ and\ \bibinfo {author}
  {\bibfnamefont {M.~W.}\ \bibnamefont {Zwierlein}},\ }\href {\doibase
  10.1103/PhysRevLett.109.095302} {\bibfield  {journal} {\bibinfo  {journal}
  {Phys. Rev. Lett.}\ }\textbf {\bibinfo {volume} {109}},\ \bibinfo {pages}
  {095302} (\bibinfo {year} {2012})}\BibitemShut {NoStop}%
\bibitem [{\citenamefont {Galitski}\ and\ \citenamefont
  {Spielman}(2013)}]{2013_Galitski_NatReview}%
  \BibitemOpen
  \bibfield  {author} {\bibinfo {author} {\bibfnamefont {V.}~\bibnamefont
  {Galitski}}\ and\ \bibinfo {author} {\bibfnamefont {I.~B.}\ \bibnamefont
  {Spielman}},\ }\href {http://dx.doi.org/10.1038/nature11841} {\bibfield
  {journal} {\bibinfo  {journal} {Nature}\ }\textbf {\bibinfo {volume} {494}},\
  \bibinfo {pages} {49} (\bibinfo {year} {2013})}\BibitemShut {NoStop}%
\bibitem [{\citenamefont {Xu}\ \emph {et~al.}(2013)\citenamefont {Xu},
  \citenamefont {You},\ and\ \citenamefont {Ueda}}]{2013_Xu_You_PRA}%
  \BibitemOpen
  \bibfield  {author} {\bibinfo {author} {\bibfnamefont {Z.-F.}\ \bibnamefont
  {Xu}}, \bibinfo {author} {\bibfnamefont {L.}~\bibnamefont {You}}, \ and\
  \bibinfo {author} {\bibfnamefont {M.}~\bibnamefont {Ueda}},\ }\href {\doibase
  10.1103/PhysRevA.87.063634} {\bibfield  {journal} {\bibinfo  {journal} {Phys.
  Rev. A}\ }\textbf {\bibinfo {volume} {87}},\ \bibinfo {pages} {063634}
  (\bibinfo {year} {2013})}\BibitemShut {NoStop}%
\bibitem [{\citenamefont {Anderson}\ \emph {et~al.}(2013)\citenamefont
  {Anderson}, \citenamefont {Spielman},\ and\ \citenamefont
  {Juzeli\ifmmode~\bar{u}\else \={u}\fi{}nas}}]{2013_Anderson_Spielman_PRL}%
  \BibitemOpen
  \bibfield  {author} {\bibinfo {author} {\bibfnamefont {B.~M.}\ \bibnamefont
  {Anderson}}, \bibinfo {author} {\bibfnamefont {I.~B.}\ \bibnamefont
  {Spielman}}, \ and\ \bibinfo {author} {\bibfnamefont {G.}~\bibnamefont
  {Juzeli\ifmmode~\bar{u}\else \={u}\fi{}nas}},\ }\href {\doibase
  10.1103/PhysRevLett.111.125301} {\bibfield  {journal} {\bibinfo  {journal}
  {Phys. Rev. Lett.}\ }\textbf {\bibinfo {volume} {111}},\ \bibinfo {pages}
  {125301} (\bibinfo {year} {2013})}\BibitemShut {NoStop}%
\bibitem [{\citenamefont {Zhai}(2015)}]{2015_Zhai_RPP}%
  \BibitemOpen
  \bibfield  {author} {\bibinfo {author} {\bibfnamefont {H.}~\bibnamefont
  {Zhai}},\ }\href {http://stacks.iop.org/0034-4885/78/i=2/a=026001} {\bibfield
   {journal} {\bibinfo  {journal} {Rep. Prog. Phys.}\ }\textbf
  {\bibinfo {volume} {78}},\ \bibinfo {pages} {026001} (\bibinfo {year}
  {2015})}\BibitemShut {NoStop}%
\bibitem [{\citenamefont {Huang}\ \emph {et~al.}(2016)\citenamefont {Huang},
  \citenamefont {Meng}, \citenamefont {Wang}, \citenamefont {Peng},
  \citenamefont {Zhang}, \citenamefont {Chen}, \citenamefont {Li},
  \citenamefont {Zhou},\ and\ \citenamefont
  {Zhang}}]{2016_Huang_Zhang_NatPhys}%
  \BibitemOpen
  \bibfield  {author} {\bibinfo {author} {\bibfnamefont {L.}~\bibnamefont
  {Huang}}, \bibinfo {author} {\bibfnamefont {Z.}~\bibnamefont {Meng}},
  \bibinfo {author} {\bibfnamefont {P.}~\bibnamefont {Wang}}, \bibinfo {author}
  {\bibfnamefont {P.}~\bibnamefont {Peng}}, \bibinfo {author} {\bibfnamefont
  {S.-L.}\ \bibnamefont {Zhang}}, \bibinfo {author} {\bibfnamefont
  {L.}~\bibnamefont {Chen}}, \bibinfo {author} {\bibfnamefont {D.}~\bibnamefont
  {Li}}, \bibinfo {author} {\bibfnamefont {Q.}~\bibnamefont {Zhou}}, \ and\
  \bibinfo {author} {\bibfnamefont {J.}~\bibnamefont {Zhang}},\ }\href
  {http://www.nature.com/nphys/journal/v12/n6/abs/nphys3672.html} {\bibfield
  {journal} {\bibinfo  {journal} {Nature Physics}\ }\textbf {\bibinfo {volume}
  {12}},\ \bibinfo {pages} {540} (\bibinfo {year} {2016})}\BibitemShut
  {NoStop}%
\bibitem [{\citenamefont {Wu}\ \emph {et~al.}(2016)\citenamefont {Wu},
  \citenamefont {Zhang}, \citenamefont {Sun}, \citenamefont {Xu}, \citenamefont
  {Wang}, \citenamefont {Ji}, \citenamefont {Deng}, \citenamefont {Chen},
  \citenamefont {Liu},\ and\ \citenamefont {Pan}}]{2016_Wu_Pan_Science}%
  \BibitemOpen
  \bibfield  {author} {\bibinfo {author} {\bibfnamefont {Z.}~\bibnamefont
  {Wu}}, \bibinfo {author} {\bibfnamefont {L.}~\bibnamefont {Zhang}}, \bibinfo
  {author} {\bibfnamefont {W.}~\bibnamefont {Sun}}, \bibinfo {author}
  {\bibfnamefont {X.-T.}\ \bibnamefont {Xu}}, \bibinfo {author} {\bibfnamefont
  {B.-Z.}\ \bibnamefont {Wang}}, \bibinfo {author} {\bibfnamefont {S.-C.}\
  \bibnamefont {Ji}}, \bibinfo {author} {\bibfnamefont {Y.}~\bibnamefont
  {Deng}}, \bibinfo {author} {\bibfnamefont {S.}~\bibnamefont {Chen}}, \bibinfo
  {author} {\bibfnamefont {X.-J.}\ \bibnamefont {Liu}}, \ and\ \bibinfo
  {author} {\bibfnamefont {J.-W.}\ \bibnamefont {Pan}},\ }\href {\doibase
  10.1126/science.aaf6689} {\bibfield  {journal} {\bibinfo  {journal}
  {Science}\ }\textbf {\bibinfo {volume} {354}},\ \bibinfo {pages} {83}
  (\bibinfo {year} {2016})}\BibitemShut {NoStop}%
\bibitem [{\citenamefont {Grusdt}\ \emph {et~al.}(2017)\citenamefont {Grusdt},
  \citenamefont {Li}, \citenamefont {Bloch},\ and\ \citenamefont
  {Demler}}]{2017_Bloch_SOC}%
  \BibitemOpen
  \bibfield  {author} {\bibinfo {author} {\bibfnamefont {F.}~\bibnamefont
  {Grusdt}}, \bibinfo {author} {\bibfnamefont {T.}~\bibnamefont {Li}}, \bibinfo
  {author} {\bibfnamefont {I.}~\bibnamefont {Bloch}}, \ and\ \bibinfo {author}
  {\bibfnamefont {E.}~\bibnamefont {Demler}},\ }\href {\doibase
  10.1103/PhysRevA.95.063617} {\bibfield  {journal} {\bibinfo  {journal} {Phys.
  Rev. A}\ }\textbf {\bibinfo {volume} {95}},\ \bibinfo {pages} {063617}
  (\bibinfo {year} {2017})}\BibitemShut {NoStop}%
\bibitem [{\citenamefont {Hasan}\ and\ \citenamefont
  {Kane}(2010)}]{2010_Hasan_RMP}%
  \BibitemOpen
  \bibfield  {author} {\bibinfo {author} {\bibfnamefont {M.~Z.}\ \bibnamefont
  {Hasan}}\ and\ \bibinfo {author} {\bibfnamefont {C.~L.}\ \bibnamefont
  {Kane}},\ }\href {\doibase 10.1103/RevModPhys.82.3045} {\bibfield  {journal}
  {\bibinfo  {journal} {Rev. Mod. Phys.}\ }\textbf {\bibinfo {volume} {82}},\
  \bibinfo {pages} {3045} (\bibinfo {year} {2010})}\BibitemShut {NoStop}%
\bibitem [{\citenamefont {Chiu}\ \emph {et~al.}(2016)\citenamefont {Chiu},
  \citenamefont {Teo}, \citenamefont {Schnyder},\ and\ \citenamefont
  {Ryu}}]{2016_Chiu_RMP}%
  \BibitemOpen
  \bibfield  {author} {\bibinfo {author} {\bibfnamefont {C.-K.}\ \bibnamefont
  {Chiu}}, \bibinfo {author} {\bibfnamefont {J.~C.~Y.}\ \bibnamefont {Teo}},
  \bibinfo {author} {\bibfnamefont {A.~P.}\ \bibnamefont {Schnyder}}, \ and\
  \bibinfo {author} {\bibfnamefont {S.}~\bibnamefont {Ryu}},\ }\href {\doibase
  10.1103/RevModPhys.88.035005} {\bibfield  {journal} {\bibinfo  {journal}
  {Rev. Mod. Phys.}\ }\textbf {\bibinfo {volume} {88}},\ \bibinfo {pages}
  {035005} (\bibinfo {year} {2016})}\BibitemShut {NoStop}%
\bibitem [{\citenamefont {Ji}\ \emph {et~al.}(2014)\citenamefont {Ji},
  \citenamefont {Zhang}, \citenamefont {Zhang}, \citenamefont {Du},
  \citenamefont {Zheng}, \citenamefont {Deng}, \citenamefont {Zhai},
  \citenamefont {Chen},\ and\ \citenamefont {Pan}}]{2014_Ji_Pan_NatPhys}%
  \BibitemOpen
  \bibfield  {author} {\bibinfo {author} {\bibfnamefont {S.-C.}\ \bibnamefont
  {Ji}}, \bibinfo {author} {\bibfnamefont {J.-Y.}\ \bibnamefont {Zhang}},
  \bibinfo {author} {\bibfnamefont {L.}~\bibnamefont {Zhang}}, \bibinfo
  {author} {\bibfnamefont {Z.-D.}\ \bibnamefont {Du}}, \bibinfo {author}
  {\bibfnamefont {W.}~\bibnamefont {Zheng}}, \bibinfo {author} {\bibfnamefont
  {Y.-J.}\ \bibnamefont {Deng}}, \bibinfo {author} {\bibfnamefont
  {H.}~\bibnamefont {Zhai}}, \bibinfo {author} {\bibfnamefont {S.}~\bibnamefont
  {Chen}}, \ and\ \bibinfo {author} {\bibfnamefont {J.-W.}\ \bibnamefont
  {Pan}},\ }\href {\doibase 10.1038/nphys2905} {\bibfield  {journal} {\bibinfo
  {journal} {Nat. Phys.}\ }\textbf {\bibinfo {volume} {10}},\ \bibinfo {pages}
  {314} (\bibinfo {year} {2014})}\BibitemShut {NoStop}%
\bibitem [{\citenamefont {Schweizer}\ \emph {et~al.}(2016)\citenamefont
  {Schweizer}, \citenamefont {Lohse}, \citenamefont {Citro},\ and\
  \citenamefont {Bloch}}]{2016_Bloch_Schweizer_arXiv}%
  \BibitemOpen
  \bibfield  {author} {\bibinfo {author} {\bibfnamefont {C.}~\bibnamefont
  {Schweizer}}, \bibinfo {author} {\bibfnamefont {M.}~\bibnamefont {Lohse}},
  \bibinfo {author} {\bibfnamefont {R.}~\bibnamefont {Citro}}, \ and\ \bibinfo
  {author} {\bibfnamefont {I.}~\bibnamefont {Bloch}},\ }\href {\doibase
  10.1103/PhysRevLett.117.170405} {\bibfield  {journal} {\bibinfo  {journal}
  {Phys. Rev. Lett.}\ }\textbf {\bibinfo {volume} {117}},\ \bibinfo {pages}
  {170405} (\bibinfo {year} {2016})}\BibitemShut {NoStop}%
\bibitem [{\citenamefont {Li}\ \emph {et~al.}(2014)\citenamefont {Li},
  \citenamefont {Natu}, \citenamefont {Paramekanti},\ and\ \citenamefont
  {Sarma}}]{2014_Li_NatComm}%
  \BibitemOpen
  \bibfield  {author} {\bibinfo {author} {\bibfnamefont {X.}~\bibnamefont
  {Li}}, \bibinfo {author} {\bibfnamefont {S.~S.}\ \bibnamefont {Natu}},
  \bibinfo {author} {\bibfnamefont {A.}~\bibnamefont {Paramekanti}}, \ and\
  \bibinfo {author} {\bibfnamefont {S.~D.}\ \bibnamefont {Sarma}},\ }\href
  {http://dx.doi.org/10.1038/ncomms6174} {\bibfield  {journal} {\bibinfo
  {journal} {Nat Commun}\ } \textbf {\bibinfo {volume} {5}},\ \bibinfo {pages}{5174} (\bibinfo
  {year} {2014})}\BibitemShut {NoStop}%
\bibitem [{\citenamefont {Li}\ and\ \citenamefont
  {Zhao}(2013)}]{li2013topological}%
  \BibitemOpen
  \bibfield  {author} { \bibinfo {author} {\bibfnamefont {X.}~\bibnamefont
  {Li}} ,\ \bibinfo {author} {\bibfnamefont {E.}~\bibnamefont
  {Zhao}} ,\ and\ \bibinfo {author} {\bibfnamefont {W.~V.}~\bibnamefont
  {Liu}} ,\ } \href {http://www.nature.com/articles/ncomms2523}
  {\bibfield  {journal} {\bibinfo  {journal} {Nat Commun}\ } \textbf {\bibinfo {volume} {4}},\ \bibinfo {pages}{1523} 
  (\bibinfo {year} {2013})}\BibitemShut {NoStop}%
\bibitem [{\citenamefont {Pethick}\ and\ \citenamefont
  {Smith}(2002)}]{pethick2002bose}%
  \BibitemOpen
  \bibfield  {author} {\bibinfo {author} {\bibfnamefont {C.~J.}\ \bibnamefont
  {Pethick}}\ and\ \bibinfo {author} {\bibfnamefont {H.}~\bibnamefont
  {Smith}},\ }\href@noop {} {\emph {\bibinfo {title} {Bose-Einstein
  condensation in dilute gases}}}\ (\bibinfo  {publisher} {Cambridge university
  press},\ \bibinfo {year} {2002})\BibitemShut {NoStop}%
\bibitem [{\citenamefont {White}(1992)}]{White1992}%
  \BibitemOpen
  \bibfield  {author} {\bibinfo {author} {\bibfnamefont {S.~R.}\ \bibnamefont
  {White}},\ }\href {https://doi.org/10.1103/PhysRevLett.69.2863} {\bibfield
  {journal} {\bibinfo  {journal} {Phys. Rev. Lett.}\ }\textbf {\bibinfo
  {volume} {69}},\ \bibinfo {pages} {2863} (\bibinfo {year}
  {1992})}\BibitemShut {NoStop}%
\bibitem [{\citenamefont {Schollw{\"o}ck}(2011)}]{Schollwock2011}%
  \BibitemOpen
  \bibfield  {author} {\bibinfo {author} {\bibfnamefont {U.}~\bibnamefont
  {Schollw{\"o}ck}},\ }\href {https://doi.org/10.1016/j.aop.2010.09.012}
  {\bibfield  {journal} {\bibinfo  {journal} {Ann. Phys.}\ }\textbf {\bibinfo
  {volume} {326}},\ \bibinfo {pages} {96} (\bibinfo {year} {2011})}\BibitemShut
  {NoStop}%
\bibitem [{\citenamefont {Frahm}\ and\ \citenamefont
  {Korepin}(1990)}]{1990_Frahm-Korepin-PRB}%
  \BibitemOpen
  \bibfield  {author} {\bibinfo {author} {\bibfnamefont {H.}~\bibnamefont
  {Frahm}}\ and\ \bibinfo {author} {\bibfnamefont {V.~E.}\ \bibnamefont
  {Korepin}},\ }\href {\doibase 10.1103/PhysRevB.42.10553} {\bibfield
  {journal} {\bibinfo  {journal} {Phys. Rev. B}\ }\textbf {\bibinfo {volume}
  {42}},\ \bibinfo {pages} {10553} (\bibinfo {year} {1990})}\BibitemShut
  {NoStop}%
\bibitem [{\citenamefont {Kuklov}\ \emph {et~al.}(2008)\citenamefont {Kuklov},
  \citenamefont {Prokof'ev}, \ and\ \citenamefont
  {Svistunov}}]{2004_Kuklov_PRL}%
  \BibitemOpen
  \bibfield  {author} {\bibinfo {author} {\bibfnamefont {A.}~\bibnamefont
  {Kuklov}}, \bibinfo {author} {\bibfnamefont {N.}\ \bibnamefont {Prokof'ev}}, \ and\
  \bibinfo {author} {\bibfnamefont {B.}\ \bibnamefont {Svistunov}},\ }\href
  {\doibase 10.1103/PhysRevLett.92.050402} {\bibfield  {journal} {\bibinfo
  {journal} {Phys. Rev. Lett.}\ }\textbf {\bibinfo {volume} {92}},\ \bibinfo
  {pages} {050402} (\bibinfo {year} {2004})}\BibitemShut {NoStop}%
\bibitem [{\citenamefont {Calabrese}\ and\ \citenamefont
  {Cardy}(2004)}]{2004_Calabrese-Cardy}%
  \BibitemOpen
  \bibfield  {author} {\bibinfo {author} {\bibfnamefont {P.}~\bibnamefont
  {Calabrese}}\ and\ \bibinfo {author} {\bibfnamefont {J.}~\bibnamefont
  {Cardy}},\ }\href {http://stacks.iop.org/1742-5468/2004/i=06/a=P06002}
  {\bibfield  {journal} {\bibinfo  {journal} {J. Stat. Mech.}\ } \bibinfo
  {pages} {P06002} (\bibinfo {year} {2004})}\BibitemShut {NoStop}%
\bibitem [{\citenamefont {Pichler}\ \emph {et~al.}(2015)\citenamefont
  {Pichler}, \citenamefont {Ramos}, \citenamefont {Daley},\ and\ \citenamefont
  {Zoller}}]{2015_Pichler_Zoller_PRA}%
  \BibitemOpen
  \bibfield  {author} {\bibinfo {author} {\bibfnamefont {H.}~\bibnamefont
  {Pichler}}, \bibinfo {author} {\bibfnamefont {T.}~\bibnamefont {Ramos}},
  \bibinfo {author} {\bibfnamefont {A.~J.}\ \bibnamefont {Daley}}, \ and\
  \bibinfo {author} {\bibfnamefont {P.}~\bibnamefont {Zoller}},\ }\href
  {\doibase 10.1103/PhysRevA.91.042116} {\bibfield  {journal} {\bibinfo
  {journal} {Phys. Rev. A}\ }\textbf {\bibinfo {volume} {91}},\ \bibinfo
  {pages} {042116} (\bibinfo {year} {2015})}\BibitemShut {NoStop}%
\bibitem [{\citenamefont {Raghu}\ \emph {et~al.}(2008)\citenamefont {Raghu},
  \citenamefont {Qi}, \citenamefont {Honerkamp},\ and\ \citenamefont
  {Zhang}}]{2008_Raghu_Qi_PRL}%
  \BibitemOpen
  \bibfield  {author} {\bibinfo {author} {\bibfnamefont {S.}~\bibnamefont
  {Raghu}}, \bibinfo {author} {\bibfnamefont {X.-L.}\ \bibnamefont {Qi}},
  \bibinfo {author} {\bibfnamefont {C.}~\bibnamefont {Honerkamp}}, \ and\
  \bibinfo {author} {\bibfnamefont {S.-C.}\ \bibnamefont {Zhang}},\ }\href
  {\doibase 10.1103/PhysRevLett.100.156401} {\bibfield  {journal} {\bibinfo
  {journal} {Phys. Rev. Lett.}\ }\textbf {\bibinfo {volume} {100}},\ \bibinfo
  {pages} {156401} (\bibinfo {year} {2008})}\BibitemShut {NoStop}%
\bibitem [{\citenamefont {Sun}\ \emph {et~al.}(2009)\citenamefont {Sun},
  \citenamefont {Yao}, \citenamefont {Fradkin},\ and\ \citenamefont
  {Kivelson}}]{2009_Sun_PRL}%
  \BibitemOpen
  \bibfield  {author} {\bibinfo {author} {\bibfnamefont {K.}~\bibnamefont
  {Sun}}, \bibinfo {author} {\bibfnamefont {H.}~\bibnamefont {Yao}}, \bibinfo
  {author} {\bibfnamefont {E.}~\bibnamefont {Fradkin}}, \ and\ \bibinfo
  {author} {\bibfnamefont {S.~A.}\ \bibnamefont {Kivelson}},\ }\href {\doibase
  10.1103/PhysRevLett.103.046811} {\bibfield  {journal} {\bibinfo  {journal}
  {Phys. Rev. Lett.}\ }\textbf {\bibinfo {volume} {103}},\ \bibinfo {pages}
  {046811} (\bibinfo {year} {2009})}\BibitemShut {NoStop}%
\bibitem [{\citenamefont {Li}\ and\ \citenamefont
  {Sarma}(2015)}]{2015_Li_NatComm}%
  \BibitemOpen
  \bibfield  {author} {\bibinfo {author} {\bibfnamefont {X.}~\bibnamefont
  {Li}} \bibinfo {author} and \bibinfo {author} {\bibfnamefont {S.~D.}\ \bibnamefont {Sarma}},\ }\href{https://www.nature.com/articles/ncomms8137} {\bibfield  {journal} {\bibinfo
  {journal} {Nat Commun}\ } \textbf {\bibinfo {volume} {6}},\ \bibinfo {pages}{7137} (\bibinfo
  {year} {2015})}\BibitemShut {NoStop}%
\bibitem [{\citenamefont {Xu}\ \emph {et~al.}(2015)\citenamefont {Xu},
  \citenamefont {Li}, \citenamefont {Zoller},\ and\ \citenamefont
  {Liu}}]{2015_Xu_Li_PRL}%
  \BibitemOpen
  \bibfield  {author} {\bibinfo {author} {\bibfnamefont {Z.-F.}\ \bibnamefont
  {Xu}}, \bibinfo {author} {\bibfnamefont {X.}~\bibnamefont {Li}}, \bibinfo
  {author} {\bibfnamefont {P.}~\bibnamefont {Zoller}}, \ and\ \bibinfo {author}
  {\bibfnamefont {W.~V.}\ \bibnamefont {Liu}},\ }\href {\doibase
  10.1103/PhysRevLett.114.125303} {\bibfield  {journal} {\bibinfo  {journal}
  {Phys. Rev. Lett.}\ }\textbf {\bibinfo {volume} {114}},\ \bibinfo {pages}
  {125303} (\bibinfo {year} {2015})}\BibitemShut {NoStop}%
\bibitem [{\citenamefont {Zhu}\ \emph {et~al.}(2016)\citenamefont {Zhu},
  \citenamefont {Gong},\ and\ \citenamefont {Sheng}}]{2016_Zhu_Sheng_PRB}%
  \BibitemOpen
  \bibfield  {author} {\bibinfo {author} {\bibfnamefont {W.}~\bibnamefont
  {Zhu}}, \bibinfo {author} {\bibfnamefont {S.~S.}\ \bibnamefont {Gong}}, \
  and\ \bibinfo {author} {\bibfnamefont {D.~N.}\ \bibnamefont {Sheng}},\ }\href
  {\doibase 10.1103/PhysRevB.94.035129} {\bibfield  {journal} {\bibinfo
  {journal} {Phys. Rev. B}\ }\textbf {\bibinfo {volume} {94}},\ \bibinfo
  {pages} {035129} (\bibinfo {year} {2016})}\BibitemShut {NoStop}%
\end{thebibliography}%

\end{document}